\documentstyle[psfig,epsfig,amssymb]{basi}
\begin{document}
\title[Spectral structure of $X$-shaped radio sources]
      {Spectral structure of $X$-shaped radio sources}
\author[D.V. Lal \& A.P. Rao]%
       {Dharam Vir Lal\thanks{e-mail: dharam@ncra.tifr.res.in}
       \& A. Pramesh Rao\thanks{e-mail: pramesh@ncra.tifr.res.in} \\ 
        National Centre for Radio Astrophysics
        (TIFR), Pune Univ. Campus, Ganeshkhind,
        Pune 411 007}
\date{Received 14 July 2004; accepted 3 September 2004}
\maketitle
\label{firstpage}
\begin{abstract}
Analysis of Giant Metrewave Radio Telescope low frequency data
for an $X$-shaped source, 3C~223.1 has revealed an unusual result as
discussed earlier by Lal \& Rao (2004).
The radio morphologies of it at 240 and 610 MHz show a well-defined
$X$-shaped structure with a pair of active jets along the north-south axis and a
pair of wings along the east-west axis, that pass symmetrically
through the undetected radio core.
The wings (or low surface brightness jet) seem to have flatter spectral
index with respect to the high surface brightness jet.
This clearly shows the value of mapping the sample of $X$-shaped
sources at low frequencies.
Here we present our preliminary results for two more such sources.
\end{abstract}


\section{Introduction}

\begin{figure*}
\begin{center}
\begin{tabular}{ll}
\includegraphics[width=5.4cm]{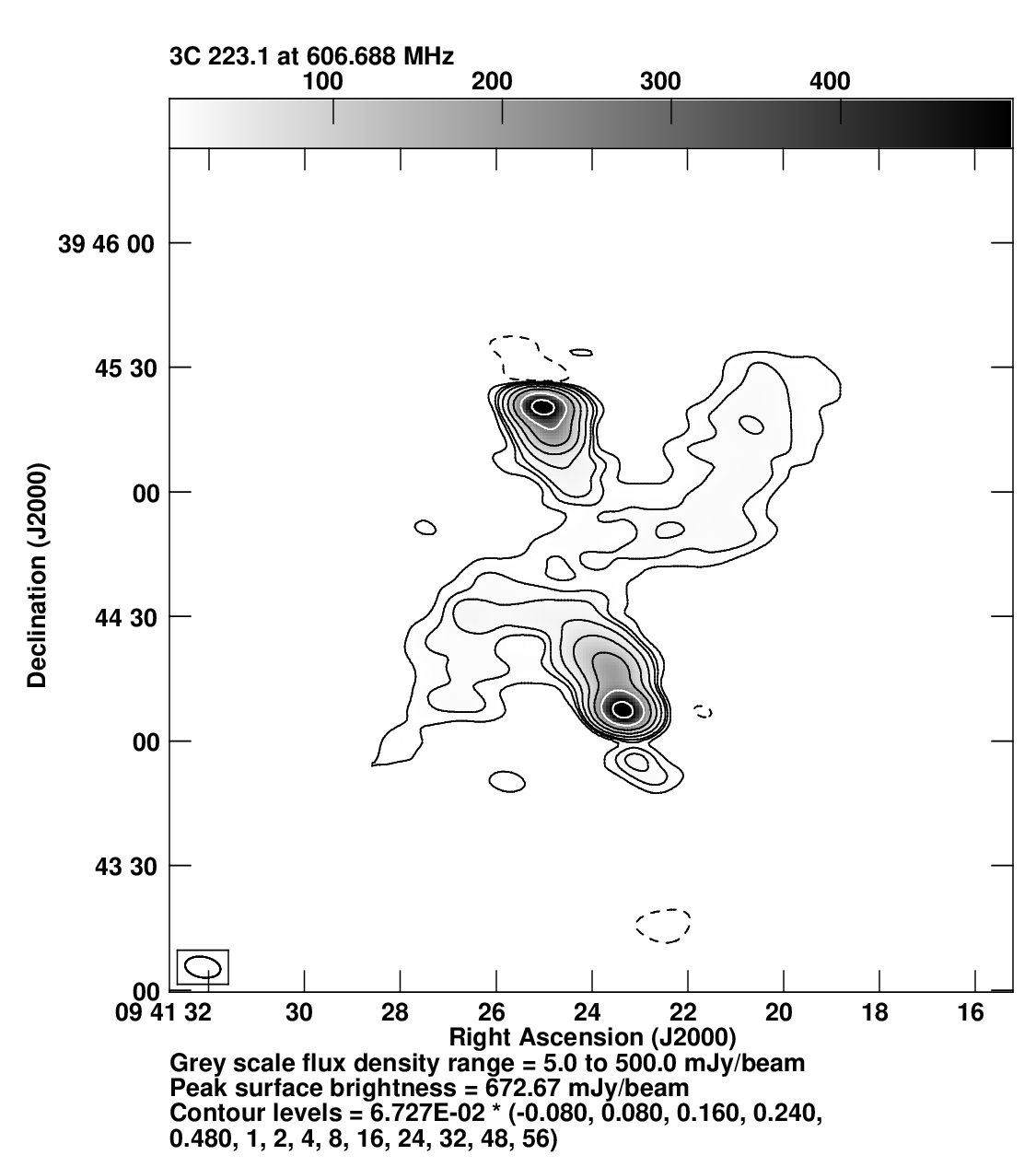} &
\includegraphics[width=5.4cm]{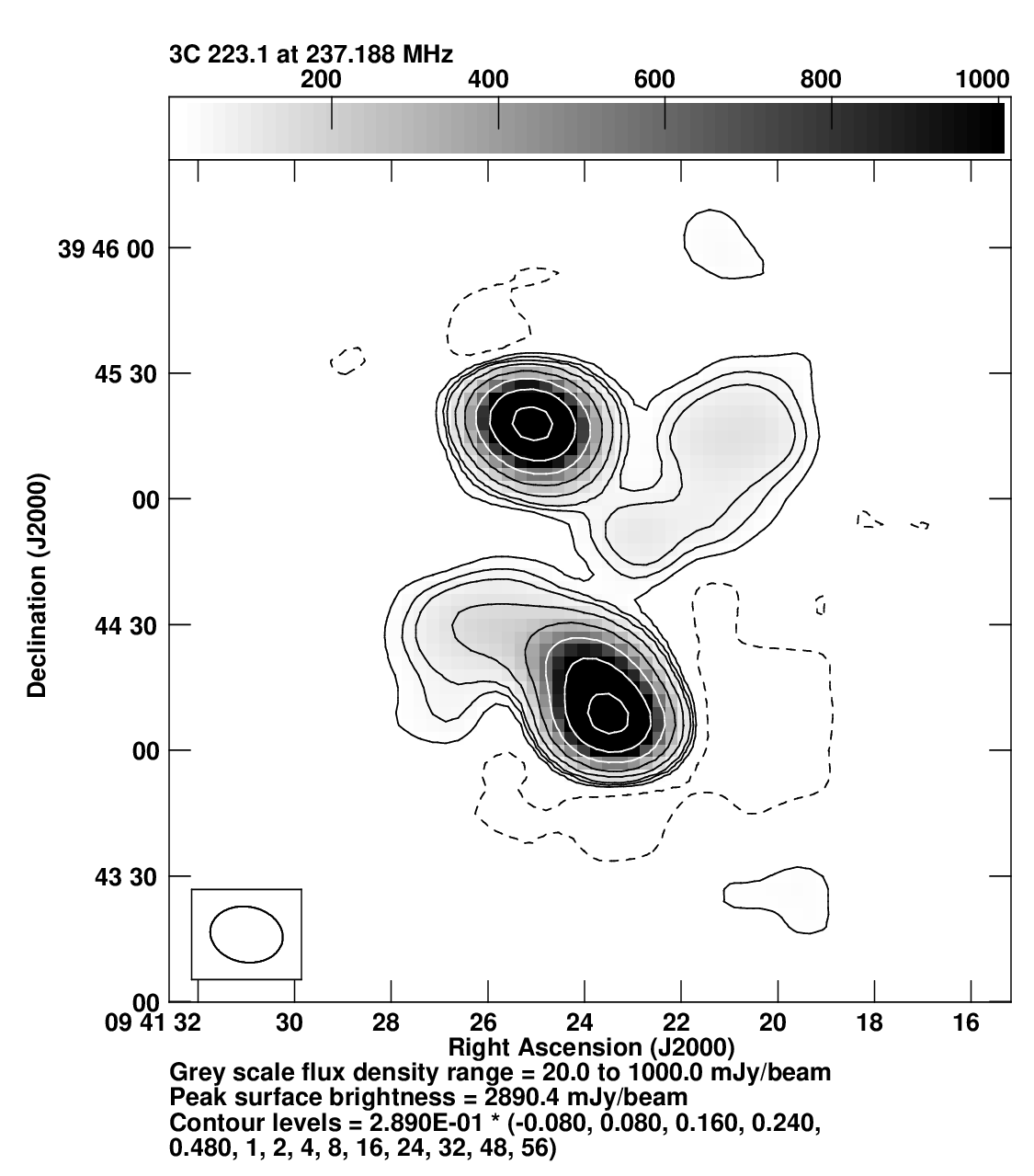} \\
\includegraphics[width=5.4cm]{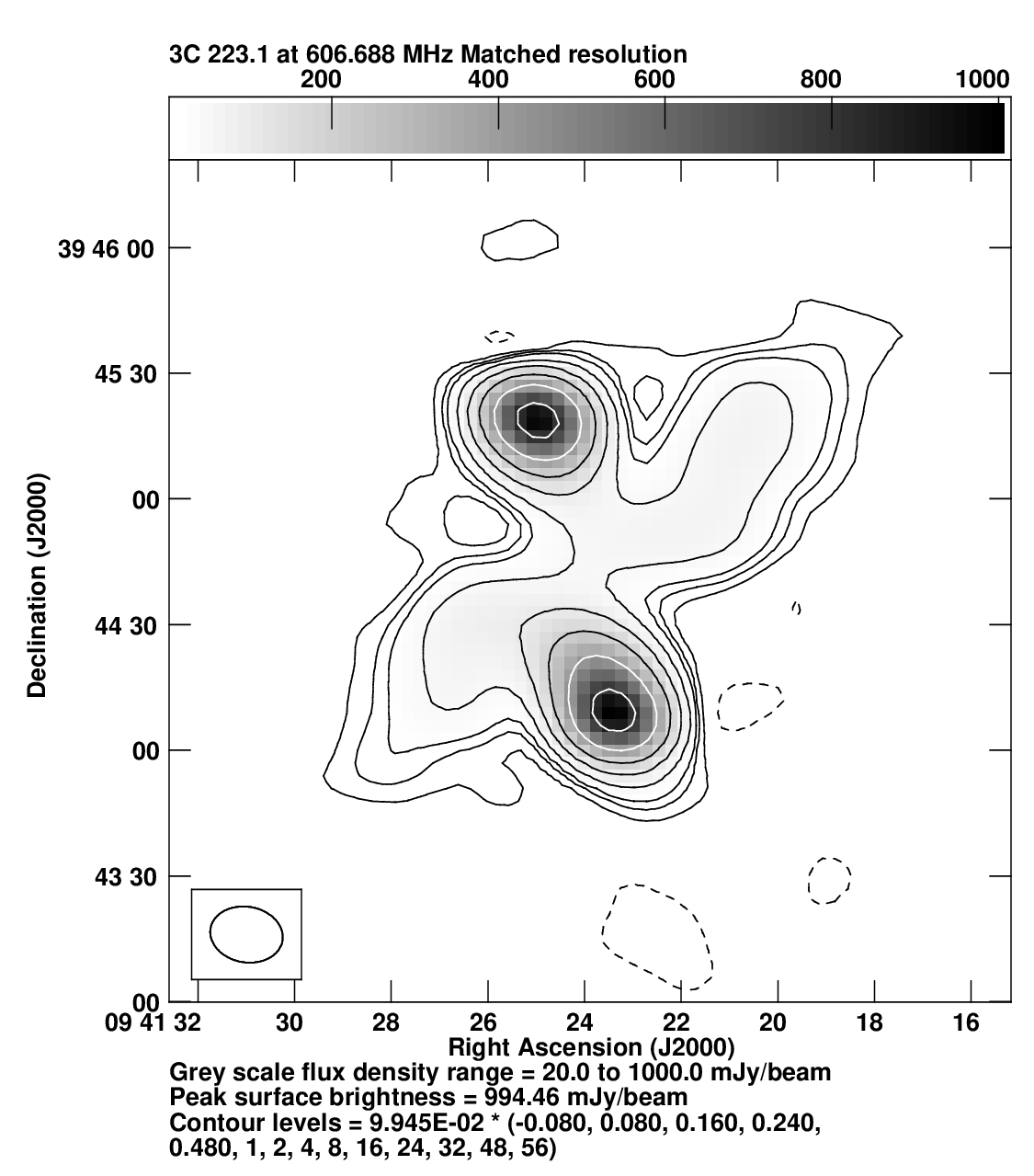} &
\includegraphics[width=5.4cm]{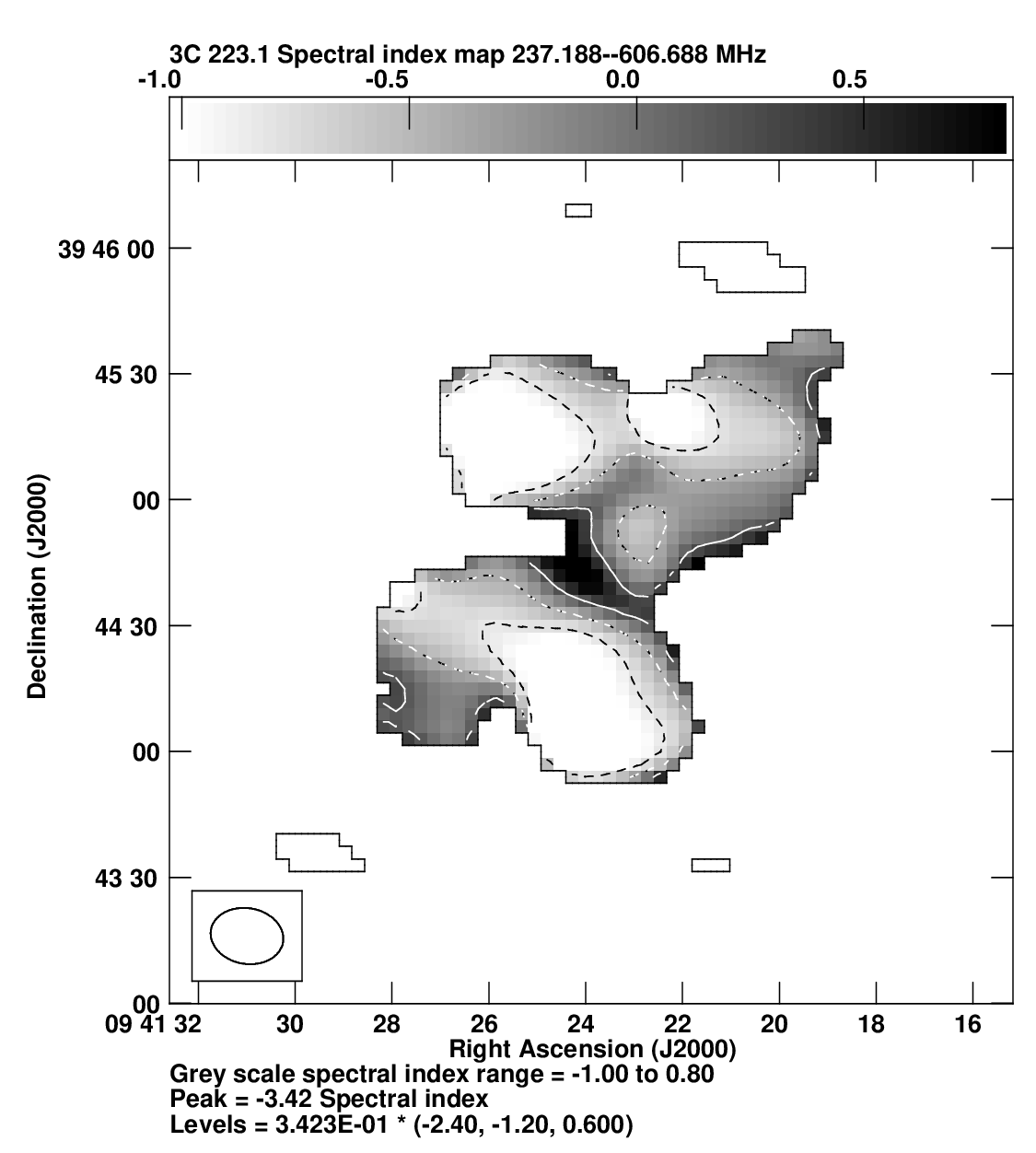}
\end{tabular}
\end{center}
\caption{Upper: Full synthesis GMRT maps of 3C 223.1 at 610
(left panel) and 240~MHz (right panel).
The CLEAN beams for 610 and 240 MHz maps are
6$^{\prime\prime}$.2~$\times$~4$^{\prime\prime}$.8
at a P.A. of 66$^{\circ}$.8
and
13$^{\prime\prime}$.1~$\times$~10$^{\prime\prime}$.8
at a P.A. of 35$^{\circ}$.2
respectively.
Lower: GMRT map of 3C 223.1 at 610~MHz
matched with the resolution of 240~MHz.
The right image is the distribution of the spectral index,
between 240 and 610 MHz, for the source.
The lighter regions represent the
relatively steep spectrum regions as compared to the darker
regions which represent flat spectrum (although the full range
of spectral index is $-$3.4 to $+$1.1, we have shown only
$-$1.0 to $+$0.8 for clarity).
}
\label{3c2231}
\end{figure*}

A class of radio sources, ``winged" or ``$X$-shaped" are
characterised by two low surface brightness lobes
(the `wings') oriented at an angle to the `active', or
high surface brightness radio lobes, giving the total source
an `$X$' shape; both sets of lobes pass
symmetrically through the centre of the associated elliptical
galaxy. Many authors have attempted to explain the unusual
structure in $X$-shaped sources. First attempt was made by
Rees (1978), who suggested that
the jet direction precesses due to a realignment caused by
the accretion of gas with respect to the central black hole
(BH) axis.
Currently, there are several models that provide different
explanations for the origin of $X$-shaped sources.
Following Dennett-Thorpe et~al. (2002), these models can
roughly be grouped into the following four formation scenarios:
($A$) Backflow from the active lobes into the wings,
($B$) slow, conical precession of the jet axis,
($C$) reorientation of the jet axis during which the flow continues and
($D$) as model $(C)$, but with the jet turned off or at
greatly reduced power during the change of direction.
Another suggested model is that of Merritt \& Ekers (2002), who identify
the winged or $X$-shaped radio sources to be the candidates
where supermassive BHs are produced by galactic mergers.
The BHs from the two galaxies fall to the centre of
the merged system and form a bound pair and the two BHs
will eventually coalesce and would lead to
reorientation of a BH spin and hence a sudden flip
in the direction of any associated jet.
A small variant of Merritt \& Ekers (2002) model was suggested by
Gopal-Krishna et~al. (2003), where these sources evolve along a $Z$-$X$
morphological sequence.
Dennett-Thorpe et~al. (2002) have argued that the wings
seen in 3C 223.1 and 3C~403 cannot
be produced by deflected backflow, and that, due to the observed
brightness of these features, buoyant expansion cannot play a large
role in their formation. A conical precession model is excluded on
morphological grounds. Their interpretation, based on the lack
of pronounced spectral gradients out to 32~GHz, even in the wings,
favours that the jet axis underwent a major reorientation in
both sources, which occurred on time-scales of no more than a few Myr.
Nevertheless, keeping in mind the {\it pros and cons} of each of these models,
most of the observational results seem to prefer possibilities
($C$) and ($D$) of Dennett-Thorpe et~al. (2002)
based on the extension of the wings and the source statistics,
and/or Merritt \& Ekers (2002).
Our recent result for 3C~223.1 (Lal \& Rao 2004,
also see Fig.~\ref{3c2231}), {\it i.e.}
the low surface brightness lobes (or the wings) have flatter
spectra as compared to high surface brightness active radio lobes,
not only does not support other less favoured models, it also 
does not support these most favoured models.
Thus, presently it is not clear as to which model
should dominate in any given source;
and more importantly, it is also not clear if we understand the
formation scenario of these $X$-shaped radio sources.

Sources B1059$+$169 (Abell~1145; redshift $z$ = 0.069) and
3C~136.1 ($z$ = 0.064)
are two of the eleven known $X$-shaped sources
(Merritt \& Ekers 2002). Source 3C~136.1
does not reside in a rich cluster, instead it probably resides in
similar environments to other `classical' Fanaroff-Riley type II sources
(FR~II, Fanaroff \& Riley 1974),
but B1059$+$169 resides in a cluster. Both sources
are of similar radio power as that of `classical' FR~IIs.
Also the host galaxy of B1059$+$169 seems to be an undisturbed
elliptical, but the host galaxy of 3C~136.1 either has a double nucleus or
is a merger remnant (Merritt \& Ekers 2002).

We have started a project to study the sample of
$X$-shaped radio sources at 240 and 610~MHz using the 
Giant Metrewave Radio Telescope (GMRT).
We have recently submitted our results for 3C~223.1
(Lal \& Rao 2004, also see Fig.~\ref{3c2231}).
Here we present preliminary results for B1059$+$169 and 3C~136.1
and describe the morphological and spectral properties.
We use the data to study  the behaviour of the
low frequency spectra~at several locations (north and south wings
and east and west lobes) across the source
and compare it with the existing models.
Our subsequent papers would discuss the radio morphologies
and statistical results that we have obtained from
radio observations using GMRT.

\section{Observations}

The 240 and 610~MHz feeds of GMRT (Ananthakrishnan \& Rao 2002) are
coaxial feeds and therefore, simultaneous multi-frequency observations
at these two frequencies are possible.
We made full-synthesis observations of B1059$+$169 and 3C~136.1 at
240 and 610 MHz, in the dual-frequency mode, using the GMRT
on 8 Feb 2004 and 9 Jan 2003 respectively, in the standard
spectral line mode with a spectral resolution of 125~kHz.
The visibility data were converted to FITS and analyzed using standard AIPS.
The flux density calibrator 3C~147 was observed
in the end for each source as an amplitude calibrator and also to estimate 
and correct for
the bandpass shape. We used the flux density scale which is an extension
of the Baars et~al. (1977) scale to low frequencies,
using the coefficients in AIPS task `SETJY'.
Sources B1021$+$219 and B0521$+$166 were used as the phase calibrator
for B1059$+$169 and 3C~136.1 respectively,
and were observed once every 35~min.
The error in the estimated flux density,
both due to calibration and systematic, is $\lesssim$ 5~\%.
The data suffered from scintillations and
intermittent radio frequency interference (RFI).
In addition to normal editing of the data, the
scintillation-affected data and
channels affected due to RFI were identified and edited,
after which the central channels were averaged using AIPS task
`SPLAT'. To avoid bandwidth smearing, for B1059$+$169 source,
6.75~MHz of clean band at 240~MHz
was reduced to 6 channels of 1.125~MHz each; and
at 610~MHz where there was little RFI, 13.5~MHz of
clean band was averaged to give 3 channels of 4.5~MHz each.
Similarly, for 3C~136.1 source,
6.75~MHz of clean band was averaged to give 6 channels
of 1.125~MHz each at 240~MHz; and
6.75~MHz of clean band was averaged to give 2 channels
of 3.375~MHz each at 610~MHz.

While imaging, 49 facets,
spread across $\sim$2$^\circ\times2^\circ$ field were used at 240~MHz and
9 facets covering slightly less
than 0.$^\circ7\times0.^\circ7$ field, were used at 610~MHz to map
each of the two fields using AIPS task `IMAGR'.
We used `uniform' weighting and the 3$-$D option for W~term
correction throughout our analysis.
The presence of a large number of point sources in the field
allowed us to do phase self-calibration to improve the image.
After 2-3 rounds of phase self-calibration, a final self-calibration
of both amplitude and phase was made to get the final image.
At each round of self-calibration, the image and the visibilities
were compared to check for the improvement in the source model.
The final maps were combined using AIPS task `FLATN' and corrected
for the primary beam of the GMRT antennas.

\begin{figure*}
\begin{center}
\begin{tabular}{ll}
\includegraphics[width=5.4cm]{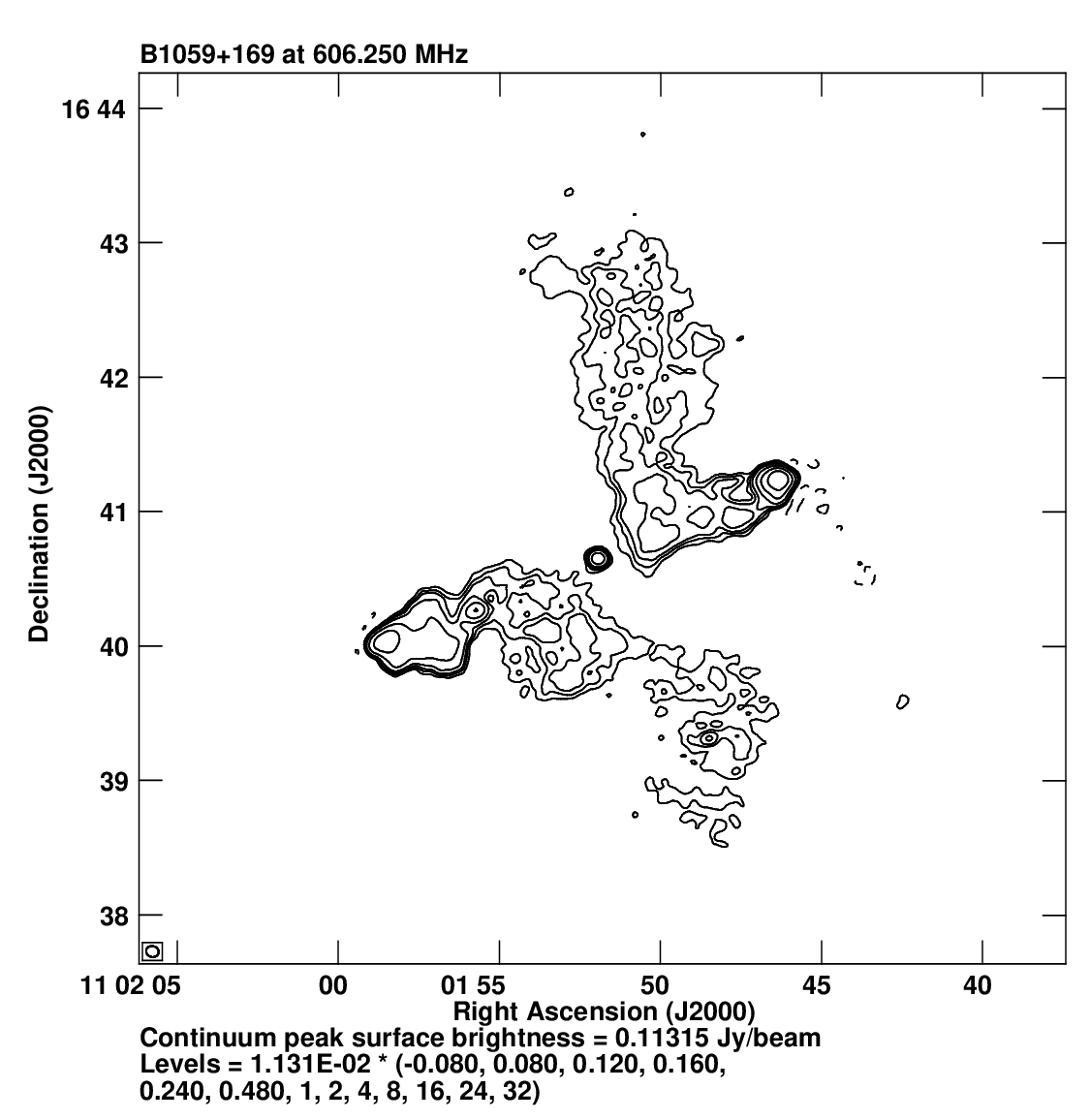} &
\includegraphics[width=5.4cm]{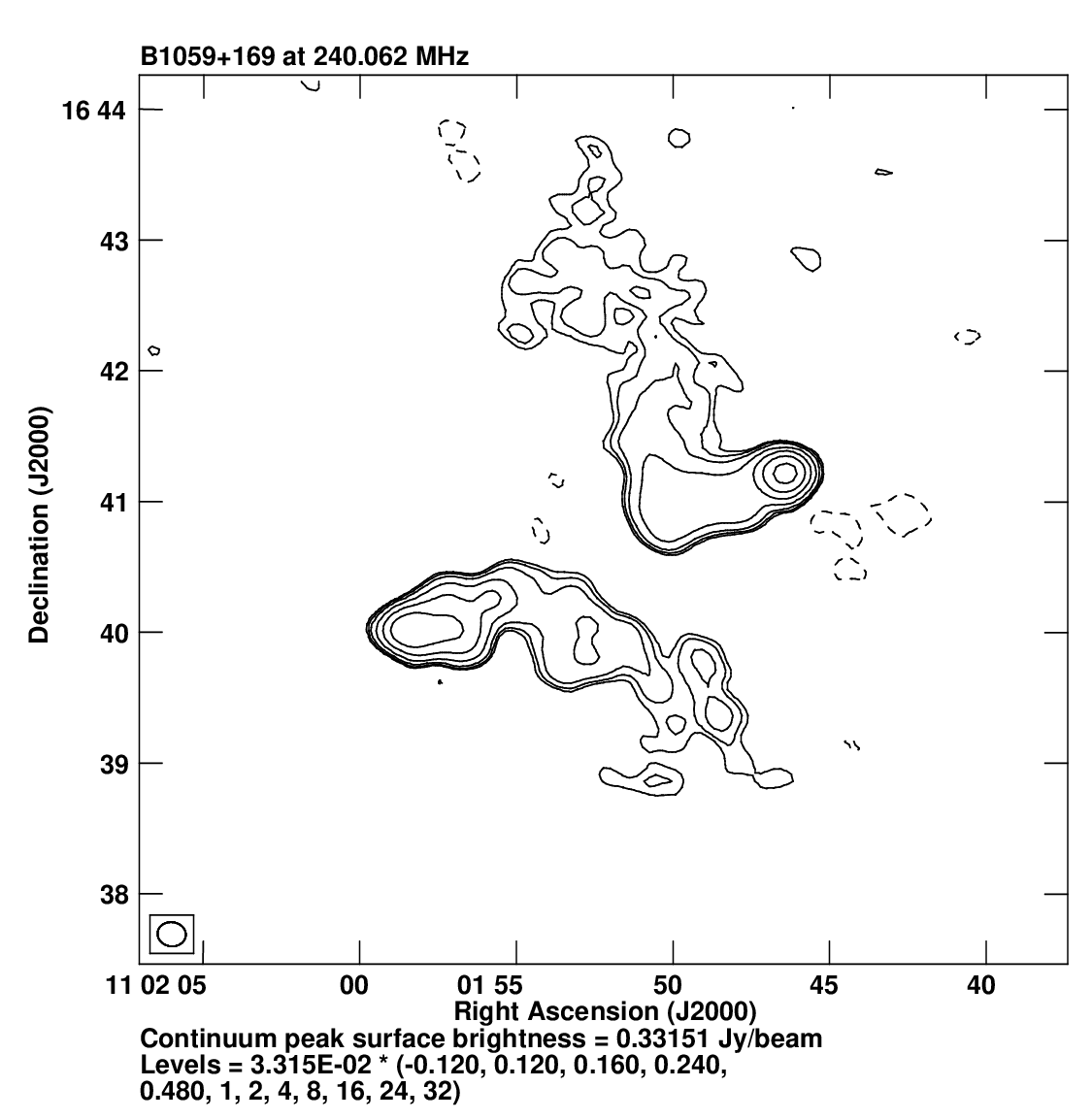} \\
\includegraphics[width=5.4cm]{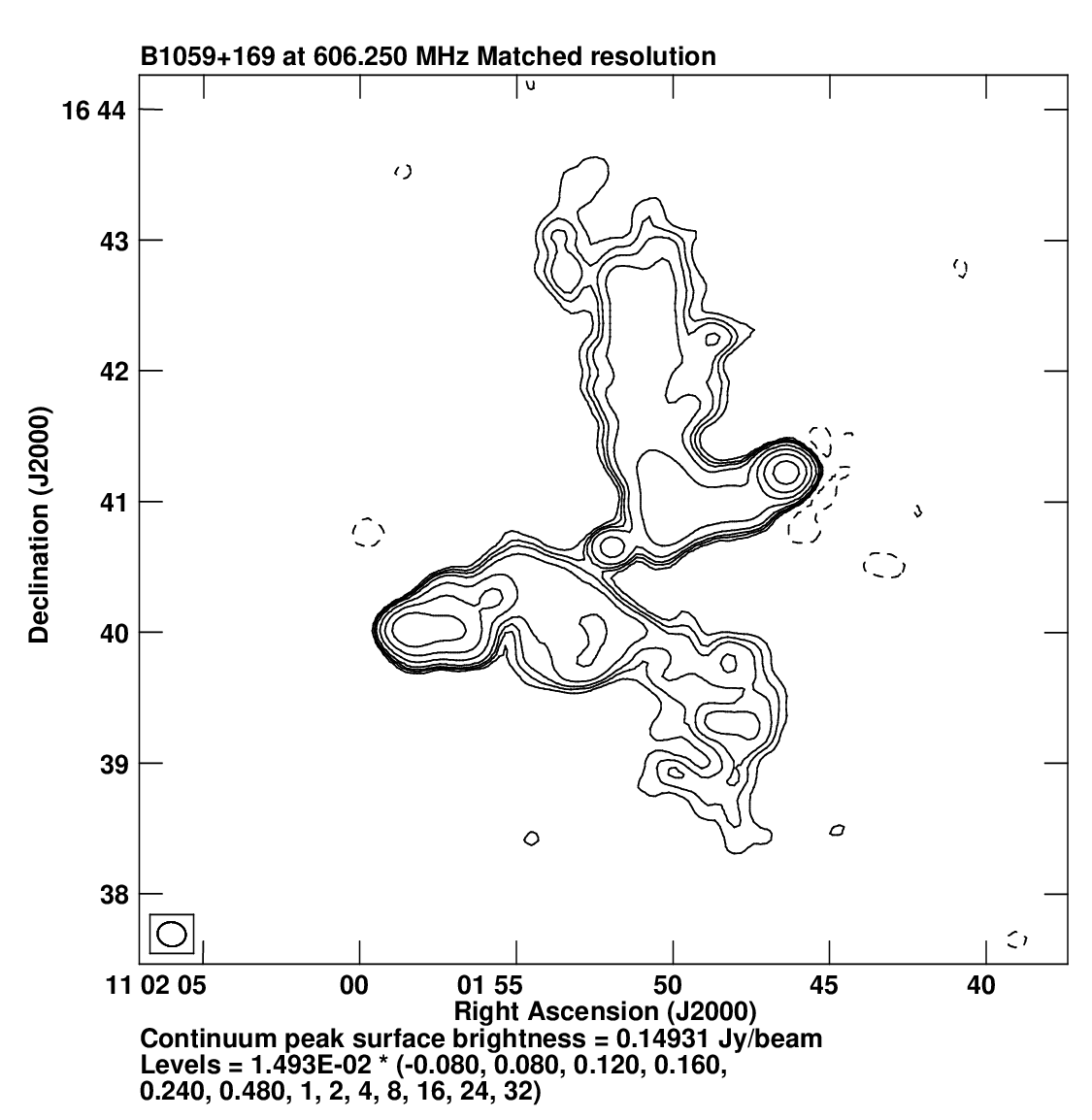} &
\includegraphics[width=5.4cm]{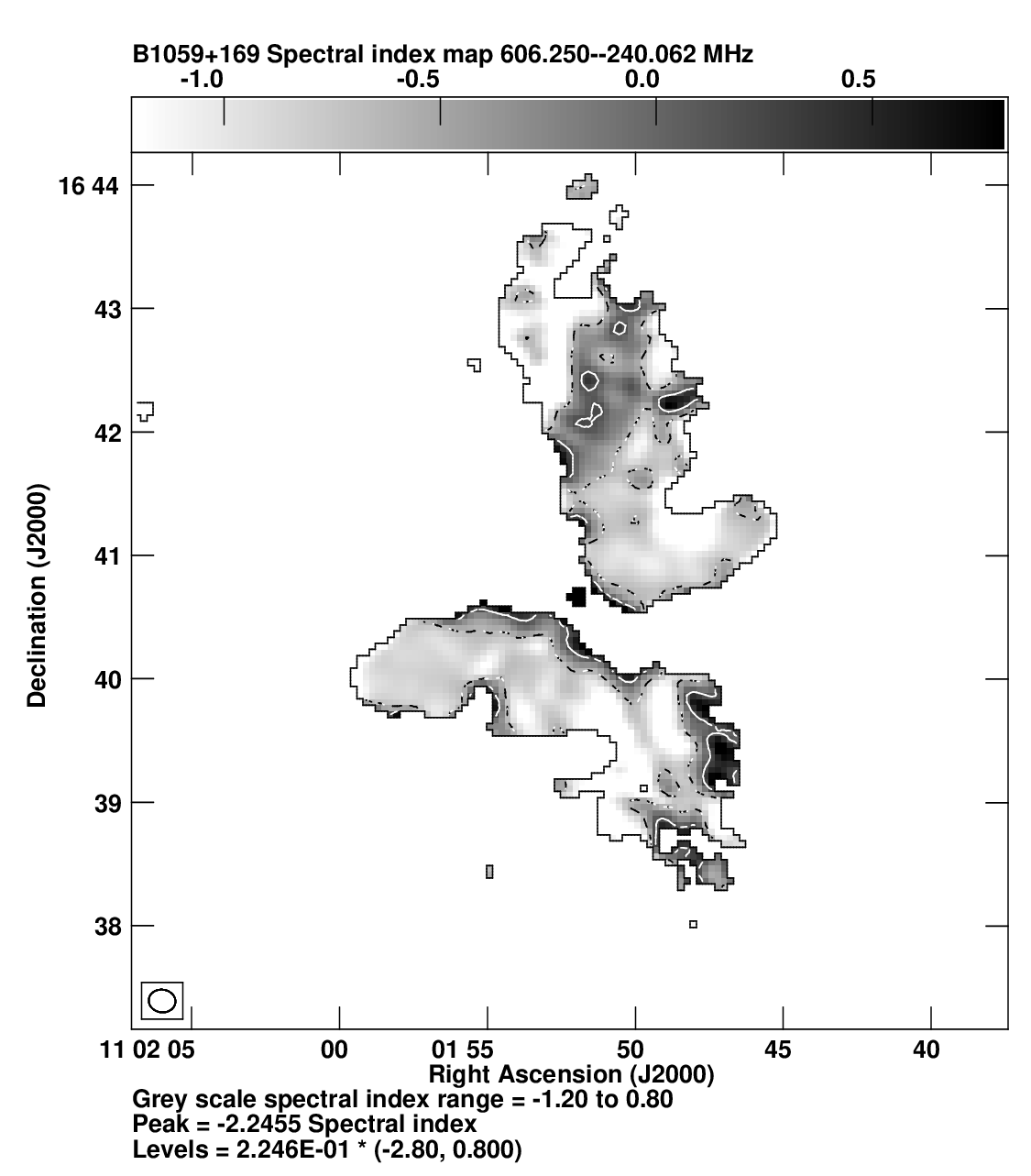}
\end{tabular}
\end{center}
\caption{Upper: Full synthesis GMRT maps of B1059$+$169 at 610
(left panel) and 240~MHz (right panel).
The CLEAN beams for 610 and 240 MHz maps are
5$^{\prime\prime}$.7~$\times$~4$^{\prime\prime}$.9
at a P.A. of 77$^{\circ}$.0
and
13$^{\prime\prime}$.0~$\times$~11$^{\prime\prime}$.1
at a P.A. of 82$^{\circ}$.5
respectively.
Lower: GMRT map of B1059$+$169 at 610~MHz
matched with the resolution of 240~MHz.
The right image is the distribution of the spectral index,
between 240 and 610 MHz, for the source.
The lighter regions represent the
relatively steep spectrum regions as compared to the darker
regions which represent flat spectrum (although the full range
of spectral index is $-$2.2 to $+$2.2, we have shown only
$-$0.8 to $+$0.8 for clarity).
}
\label{full_syn1}
\end{figure*}

\begin{figure*}
\begin{center}
\begin{tabular}{ll}
\includegraphics[width=5.4cm]{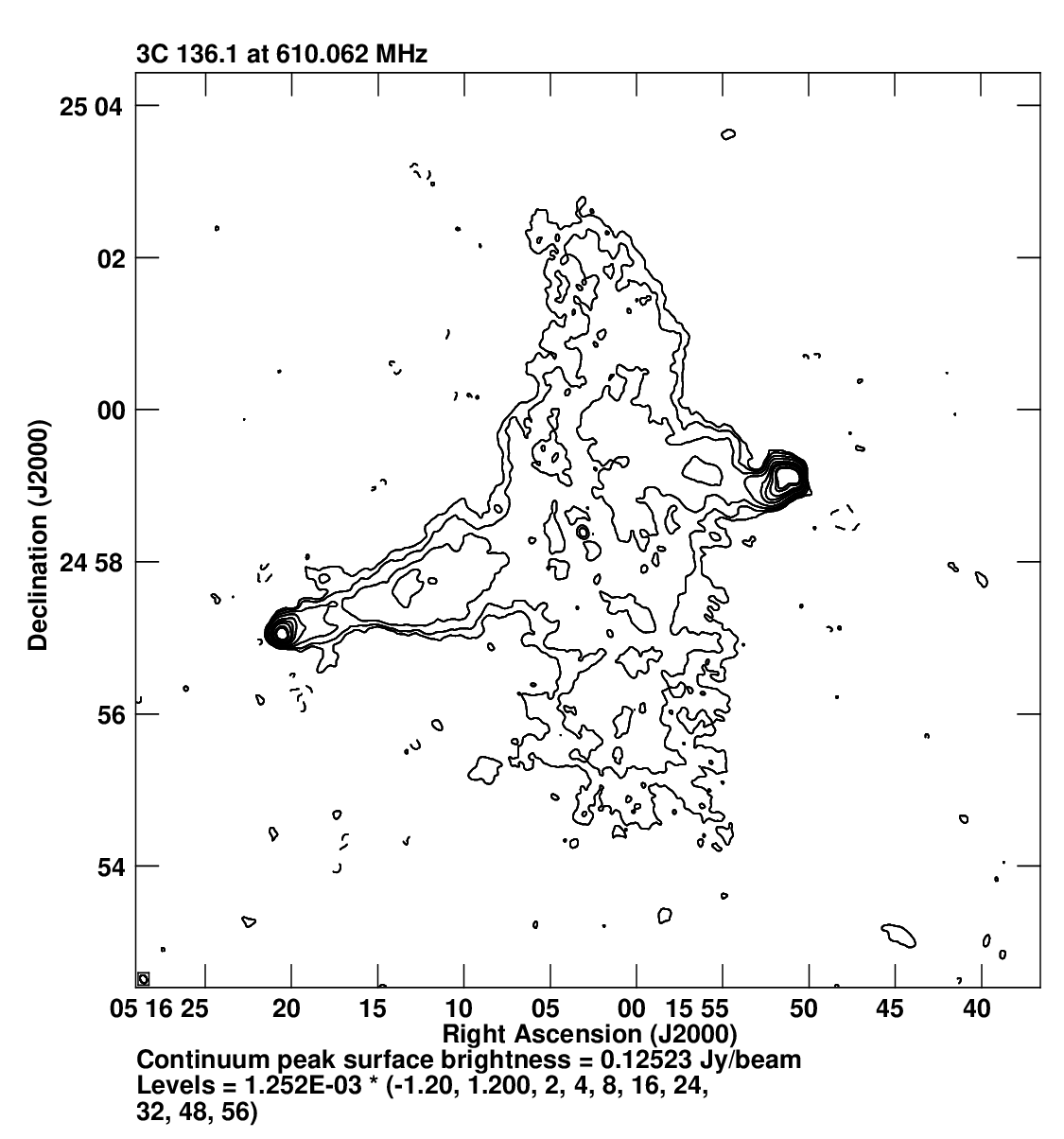} &
\includegraphics[width=5.4cm]{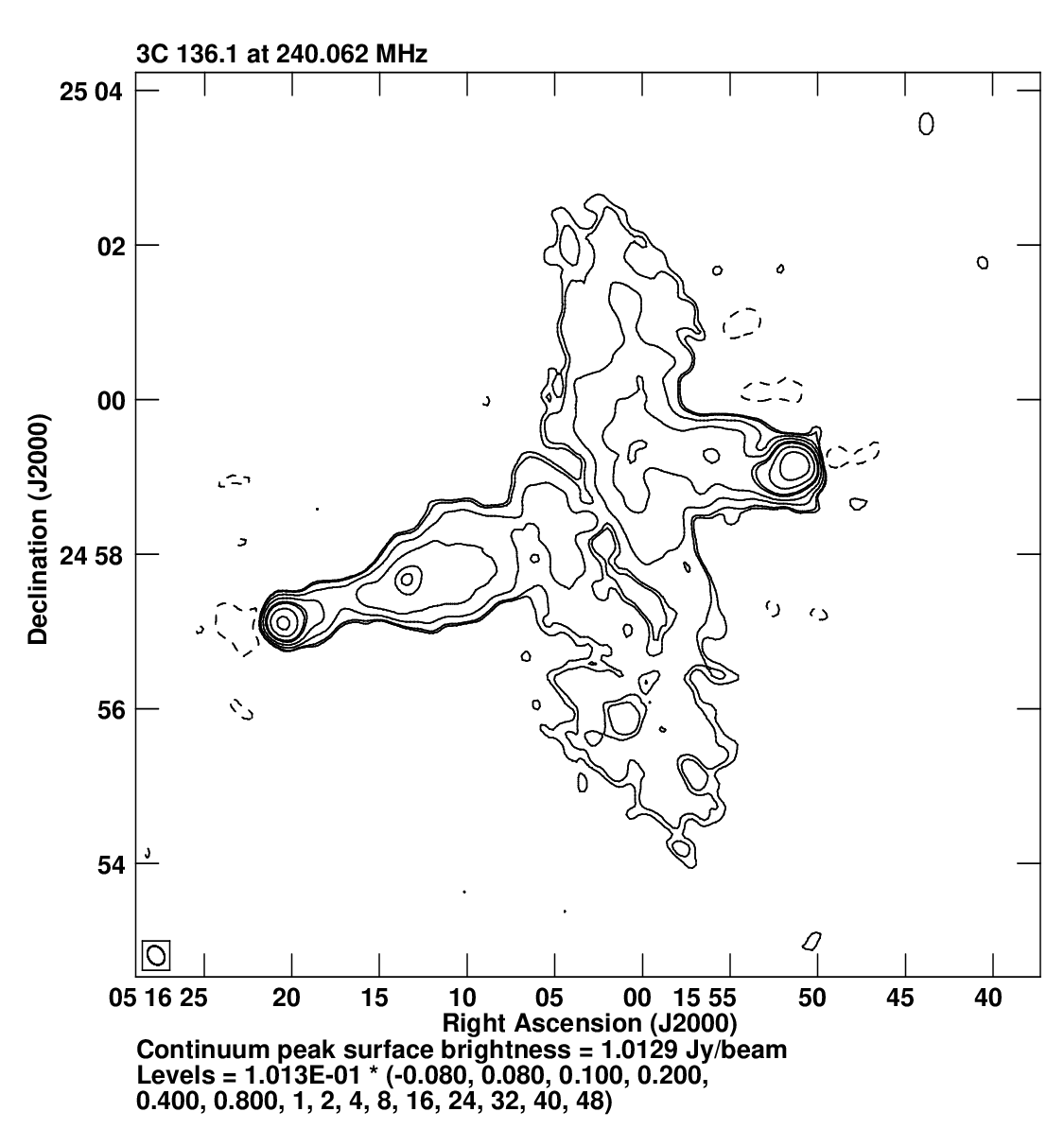} \\
\includegraphics[width=5.4cm]{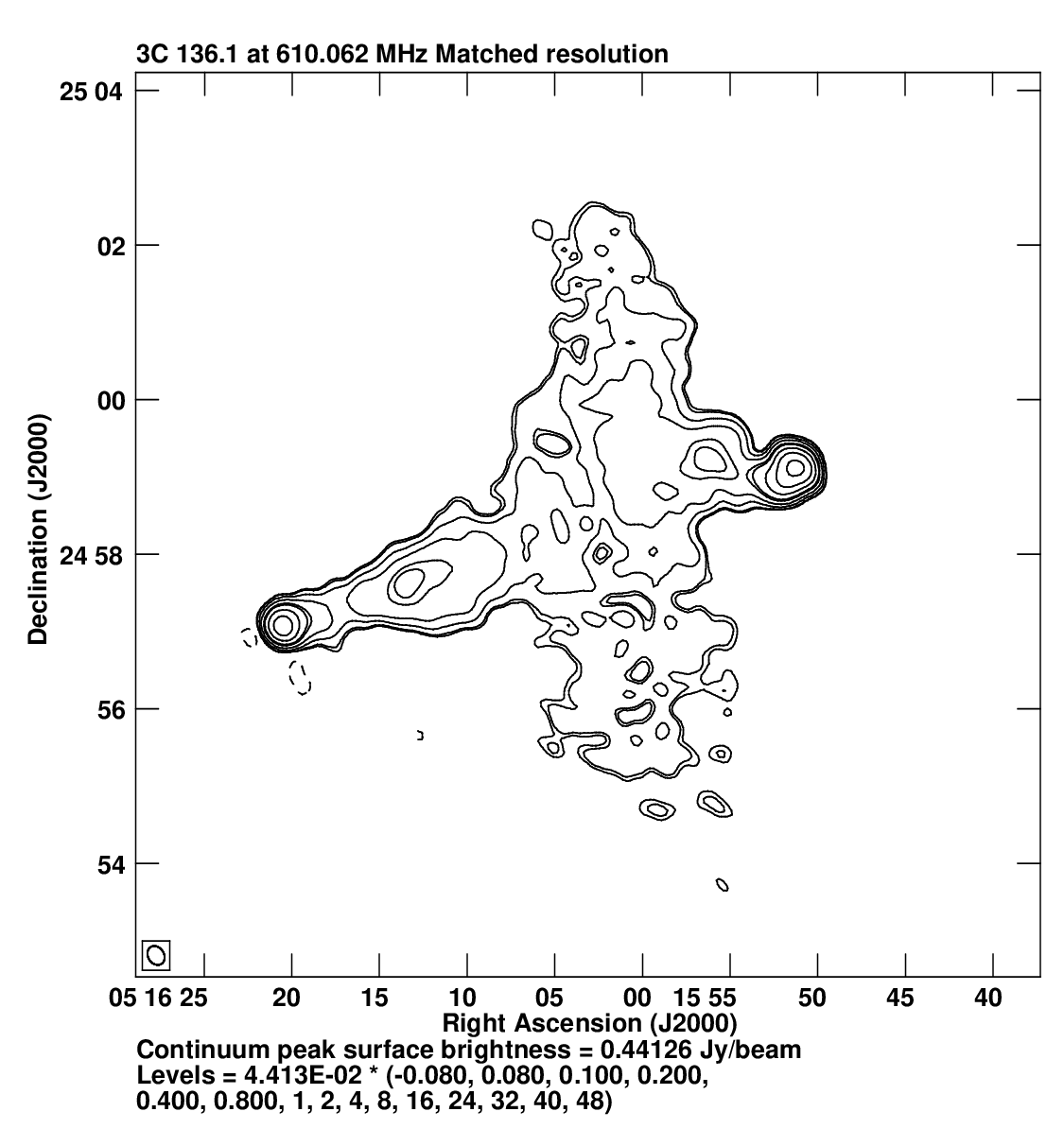} &
\includegraphics[width=5.4cm]{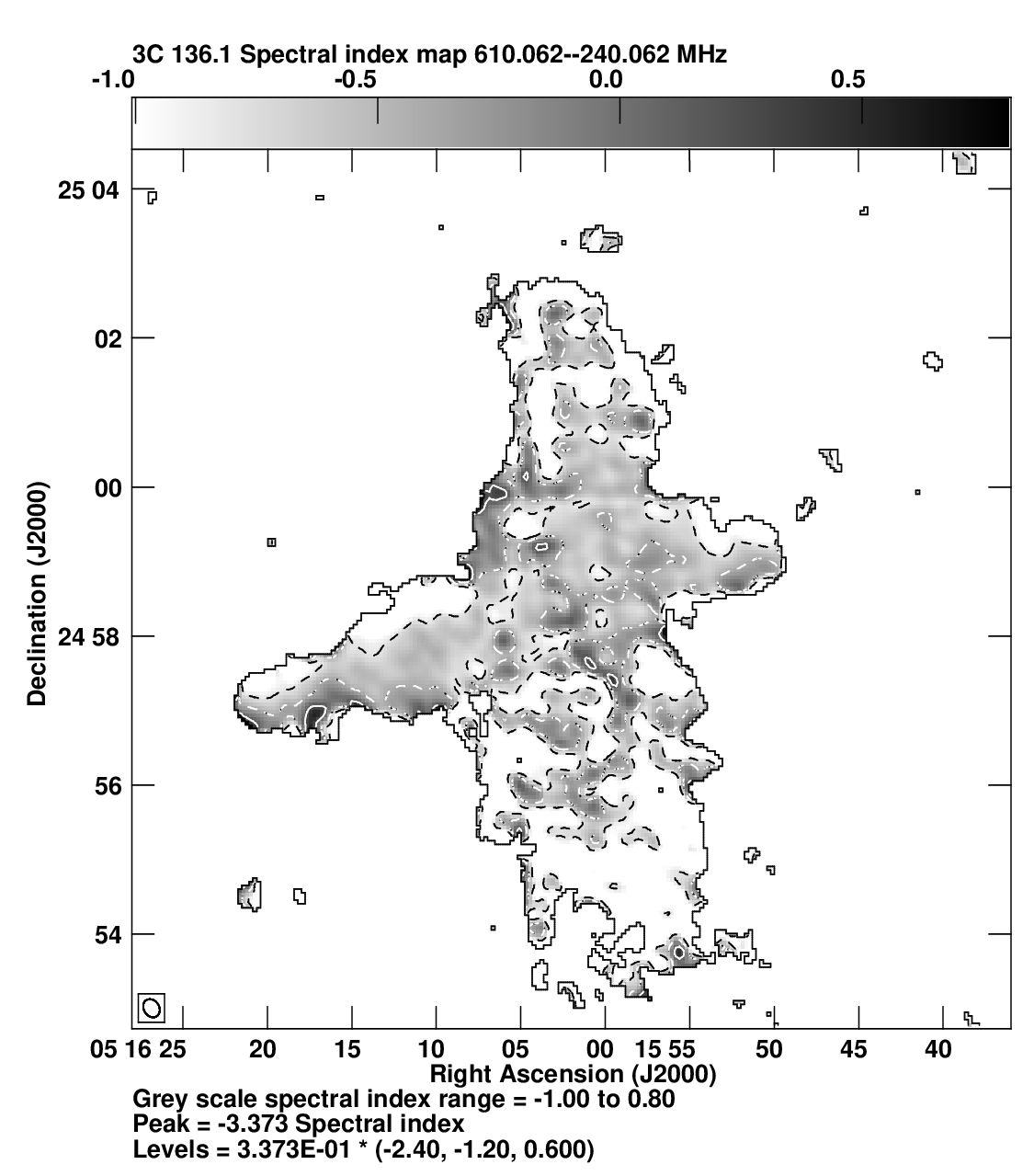} \\
\end{tabular}
\end{center}
\caption{Upper: Full synthesis GMRT maps of 3C 136.1 at 610
(left panel) and 240~MHz (right panel).
The CLEAN beams for 610 and 240 MHz maps are
6$^{\prime\prime}$.7~$\times$~5$^{\prime\prime}$.1
at a P.A. of 28$^{\circ}$.2
and
15$^{\prime\prime}$.5~$\times$~12$^{\prime\prime}$.3
at a P.A. of 32$^{\circ}$.7
respectively.
Lower: GMRT map of 3C 136.1 at 610~MHz
matched with the resolution of 240~MHz.
The right image is the distribution of the spectral index,
between 240 and 610 MHz, for the source.
The lighter regions represent the
relatively steep spectrum regions as compared to the darker
regions which represent flat spectrum (although the full range
of spectral index is $-$3.4 to $+$1.0, we have shown only
$-$1.0 to $+$0.8 for clarity).
}
\label{full_syn2}
\end{figure*}

The full synthesis radio images shown in Figs.~\ref{full_syn1}
and \ref{full_syn2}
have nearly complete UV coverage, an angular resolution
$\sim$15$^{\prime\prime}$ and $\sim$6$^{\prime\prime}$ and the
rms~noise in the maps are $\sim$2.0 and
$\sim$0.2 mJy~beam$^{-1}$ at 240 and 610~MHz respectively.
The dynamic ranges in the two maps are
$\sim$200 and $\sim$600 in case of B1059$+$169 and
$\sim$600 and $\sim$500 in case of 3C~136.1
respectively at 240 and 610~MHz.
The GMRT has a hybrid configuration (Swarup et~al. 1991;
Ananthakrishnan \& Rao 2002)
with 14 of its 30 antennas located in a central compact array
with size $\sim$1.1~km and the remaining antennas distributed in
a roughly `Y' shaped configuration, giving a maximum baseline length
of $\sim$25~km.
The hybrid configuration gives reasonably good sensitivity
for both compact and extended sources.

\section{Radio morphology}

The first high angular resolution, high sensitivity images
of 3C~223.1, B1059$+$169 and 3C~136.1 at the lowest frequencies of 240~MHz
(upper right panel) and 610~MHz (upper left panel) are shown in
Figs.~\ref{3c2231}, \ref{full_syn1} and \ref{full_syn2} respectively.

In all three cases, our estimates at both frequencies, 240 and 610~MHz
agree well with that of measurements with other instruments.
Also, the sources are small (roughly 4--8 arcmin
across) and are much smaller than the primary beam.
Furthermore, we have simulated the source at
one frequency using UV coverage corresponding to the other frequency
and {\it vice versa}; the maps show similar morphologies and flux densities.
We therefore believe that we have not
lost any flux density in our interferometric observations at
the two different frequencies.

\section{Low frequency radio spectra}

The observations and morphology described above
allow us to investigate in detail the spectral
index distributions of B1059$+$169 and 3C~136.1.
The final calibrated UV data at 610~MHz
was first mapped using UV~taper 0--22~k$\lambda$,
which is similar to that of 240~MHz data and then
restored using the restoring beam
corresponding to the 240~MHz map.
These restored and matched maps,
were used further for the spectral analysis.
We looked at the spatially `filtered' images (Wright et~al. 1999)
and `spectral tomography' images (Rudnick 2002)
to enhance weak features seen in the maps.
We do not find evidence for a new feature emerging in
the filtered or tomographic images, which could have been
missed in the single frequency maps.
While, it could also be due to sensitivity limits of our maps,
this is possibly due to the fact that these techniques
are effective in separating overlapping large-scale and
small-scale features with different spectral distributions.
Hence, these techniques are found to be useful for objects
such as radio galaxies and supernova remnants (Rudnick 2002).
We determine the spectral index distribution using
standard, direct method of determining the spectral
index between maps $S_{\nu_1}(x,y)$ and $S_{\nu_2}(x,y)$
at two frequencies ${\nu_1}$ and ${\nu_2}$. This is given by
$$
\alpha_{\nu_1,\nu_2}(x,y) \equiv
\frac{{\rm log}~(S_{\nu_1} (x,y)/S_{\nu_2} (x,y))}{{\rm log}~(\nu_1/\nu_1)}.
$$

We deduce the flux densities at 240 and 610~MHz using images shown
in Figs.~\ref{full_syn1} and \ref{full_syn2},
which are matched to the same resolution.
The flux densities for the active lobes and the wings are
integrated over the region, which is at least 4-5 times the beam
size and  above their 3$\sigma$ contour to reduce statistical errors.
Also, analysis of the spectrum in different
regions of the source shows remarkable variation across the source
(see Figs.~\ref{full_syn1} and \ref{full_syn2}, lower right panels).

\noindent{\bf {3C~223.1}}
The low frequency (240--610~MHz) fitted spectra have
$-$0.37 $< \alpha <$ $-$1.08 for all regions across the source.
The source also
shows evidence for steeper spectra in the active lobes than in the wings.
The east and west wings have low frequency (240--610~MHz) spectral indices of
$-$0.37 $\pm$0.03 and $-$0.62 $\pm$0.02 respectively,
whereas the north and south active lobes have
values of $-$1.08 $\pm$0.01 and $-$1.08 $\pm$0.01 respectively.
These results at low frequencies are consistent with those
obtained at high frequency (Dennett-Thorpe et~al. 2002); i.e.
the high frequency (1.4--32~GHz) spectral indices were respectively
$-$0.70 $\pm$0.03 and $-$0.66 $\pm$0.03 for east and west wings, and
$-$0.75 $\pm$0.02 and $-$0.77 $\pm$0.02 for north and south active lobes.
Or, in other words, low surface brightness lobes have flatter
spectra as compared to high surface brightness active radio lobes.

\noindent{\bf {B1059$+$169}}
Here, The low frequency (240--610~MHz) fitted spectra have
$-$0.22 $< \alpha <$ $-$0.88 for all regions across the source.
The source shows evidence for steeper spectra in the active lobes
than in the wings, similar to 3C~223.1.
The east and west active lobes have low frequency (240--610~MHz)
spectral indices of
$-$0.88 $\pm$0.01 and $-$0.76 $\pm$0.01 respectively,
whereas the north and south
wings have values of $-$0.22 $\pm$0.02 and $-$0.53 $\pm$0.02 respectively.
Or, in other words, low surface brightness lobes have flatter
spectra as compared to high surface brightness active radio lobes.
 
\noindent{\bf {3C~136.1}}
Here, the low frequency (240--610~MHz) fitted spectra have
$-$0.61 $< \alpha <$ $-$0.76 for all regions across the source.
This source does not
show evidence for steeper spectra in the active lobes than in the wings.
Instead, the east and west active lobes have low frequency
(240--610~MHz) spectral indices of
$-$0.61 $\pm$0.01 and $-$0.61 $\pm$0.02 respectively,
whereas the north and south
wings have values of $-$0.76 $\pm$0.01 and $-$0.71 $\pm$0.02 respectively.
Or, in other words, it is the high surface brightness active lobes 
that have flatter spectra as compared to low surface brightness
(wings) radio lobes.

The errors in spectral index corresponding to errors in the estimated
flux density, both due to calibration and systematic, are less than
0.07. Even if we choose more conservative value of 10~\% as error in
the estimated flux density, the error in spectral index is $\sim$~0.15.
Since, the observed differences in spectral index are much more
between the wings and active lobes in 3C~223.1 and B1059$+$169,
we believe that the observed spectral index features are real.
 
\section{Results}

We have presented the low frequency images of 3C~223.1,
B1059$+$169 and 3C~136.1 at 240 and 610~MHz. We find that the low
surface brightness lobes (or the wings) for 3C~223.1 and B1059$+$169
have flatter spectra as compared to high surface brightness active
radio lobes.
In case of 3C~136.1, it is the high surface brightness active
radio lobes, which have flatter spectra as compared to the
low surface brightness lobes (or the wings).

The important consequence of our observations for $X$-shaped sources
is that there is some ambiguity as to the formation of $X$-shaped sources.
Low frequency, high resolution radio images for all the sample
$X$-shaped sources would be useful in independently establishing
the formation models for these sources, which would be discussed in
our subsequent papers.

\section*{Acknowledgments}

We thank the staff of the GMRT who have made these observations
possible. GMRT is run by the National Centre for Radio Astrophysics
of the Tata Institute of Fundamental Research.
We thank the anonymous referee for several useful comments.
DVL thanks R.~Nityananda for discussions and several useful comments.
This research has made use of the NASA/IPAC Extragalactic Database
which is operated by the Jet Propulsion Laboratory,
Caltech, under contract with NASA, and NASA's Astrophysics Data System.

\label{lastpage}
\end{document}